\documentclass[aps,pra,superscriptaddress,twocolumn]{revtex4-1}

\usepackage{graphicx}
\usepackage{amsmath}
\usepackage{amssymb}
\usepackage{hyperref}
\usepackage[utf8]{inputenc}
\usepackage{mathtools}
\usepackage[english]{babel}
\usepackage{bbm}
\hypersetup{colorlinks=true, linkcolor=blue, citecolor=blue, urlcolor=blue}
\usepackage{xcolor}
\usepackage{braket}
\usepackage[normalem]{ulem}

\newcommand{\ie}{i.\,e.,\ }

\newcommand{\eg}{e.\,g.,\ }

\newcommand{\cf}{cf.\ }
\newcommand{\re}{\mathrm{Re}}

\newcommand{\rr}{\mathbf{r}}

\newcommand{\kk}{\mathbf{k}}

\newcommand{\fref}[1]{\text{Fig.}~\ref{#1}}
\newcommand{\ffref}[1]{\text{Figs.}~\ref{#1}}
\newcommand{\eref}[1]{\text{Eq.}~\eqref{#1}}
\newcommand{\eeref}[1]{\text{Eqs.}~\eqref{#1}}
\begin{document}
\title{Superglass formation in an atomic BEC with competing long-range interactions}
\author{Stefan Ostermann}
\email{stefanostermann@g.harvard.edu}
\affiliation{Department of Physics, Harvard University, Cambridge, Massachusetts 02138, USA}
\author{Valentin Walther}
\affiliation{Department of Physics, Harvard University, Cambridge, Massachusetts 02138, USA}
\affiliation{ITAMP, Harvard-Smithsonian Center for Astrophysics, Cambridge, Massachusetts 02138, USA}
\author{Susanne F. Yelin}
\affiliation{Department of Physics, Harvard University, Cambridge, Massachusetts 02138, USA}

\begin{abstract}
The complex dynamical phases of quantum systems are dictated by atomic interactions that usually evoke an emergent periodic order. Here, we study a quantum many-body system with two competing and substantially different long-range interaction potentials where the dynamical instability towards density order can give way to a superglass phase,~\ie a superfluid disordered amorphous solid, which exhibits local density modulations but no long-range periodic order. We consider a two-dimensional BEC in the Rydberg-dressing regime coupled to an optical standing wave resonator. The dynamic pattern formation in this system is governed by the competition between the two involved interaction potentials: repulsive soft-core interactions arising due to Rydberg dressing and infinite-range sign changing interactions induced by the cavity photons. The superglass phase is found when the two interaction potentials introduce incommensurate length scales. The dynamic formation of this peculiar phase without any externally added disorder is driven by quantum fluctuations and can be attributed to frustration induced by the two competing interaction energies and length scales.
\end{abstract}

\maketitle

\section{Introduction}\label{sec:introduction}
The controllability of individual atomic systems increased tremendously over the past decades~\cite{metcalf_laser_2003, haroche_exploring_2006, bloch_many-body_2008, reiserer_cavity-based_2015, gross_quantum_2017,browaeys_many-body_2020}. This enables efficient trapping and cooling of atomic gases or individual atoms, controlling individual photons and tailoring interactions between atoms and photons. One particularly challenging and active modern research direction in this realm are systems with tailored long-range interactions~\cite{defenu_long-range_2021}. Prominent examples are, among others, dipolar Bose-Einstein condensates (BECs)~\cite{griesmaier_bose-einstein_2005, lu_strongly_2011, aikawa_bose-einstein_2012, norcia_new_2021}, ultracold atomic gases in cavities~\cite{ritsch_cold_2013, mivehvar_cavity_2021, periwal_programmable_2021} and individually trapped atoms with Rydberg interactions~\cite{ebadi_quantum_2021, scholl_quantum_2021, semeghini_probing_2021}. These systems allow for the study of artificial quantum matter in a well-controlled and tunable environment.
\begin{figure}
\centering
\includegraphics[width=0.45\textwidth]{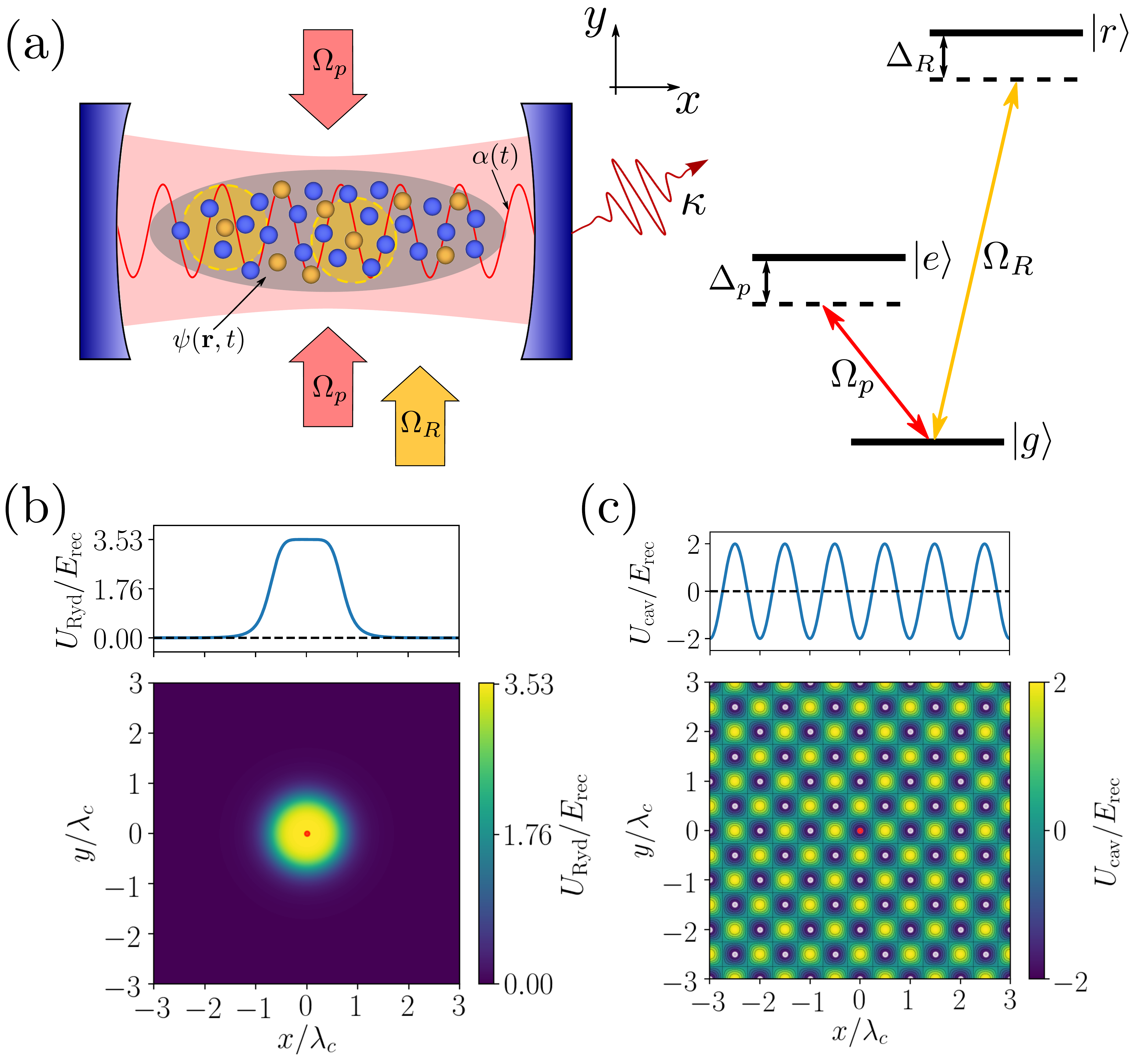}
\caption{(a) Considered setup. A Rydberg-dressed BEC is trapped and confined to two dimensions in an optical resonator with lasers impinging from the side. The BEC atoms form an effective V-level structure. The transition $\ket{g}\leftrightarrow \ket{e}$ is coupled to the cavity at a strength $\mathcal{G}_p$. The transition $\ket{g}\leftrightarrow\ket{r}$ is driven by an additional laser in the Rydberg dressing regime, implying that the high-lying Rydberg state $\ket{r}$ is only very weakly populated. The Rydberg dressing imposes a long-range soft-core interaction potential between the atoms, see panel (b). In addition, the cavity photons induce non-trivial infinite  range interactions between atoms, see panel (c). The top curves in panels (b) and (c) show cuts along the $x$-axis at $y=0.0\lambda_c$ for the two interaction potentials.}
\label{fig:setup}
\end{figure}

Over the last decade the focus of research was on investigating long-range interacting systems with crystalline properties,~\ie periodic systems with long-range order. In this realm first experimental observations of quantum phase transitions like the superfluid to Mott insulator transition~\cite{greiner_quantum_2002}, the formation of supersolid phases of matter~\cite{leonard_supersolid_2017, li_stripe_2017, tanzi_observation_2019, bottcher_transient_2019, chomaz_long-lived_2019,schuster_supersolid_2020,norcia_two-dimensional_2021,bland_two-dimensional_2021} or non-trivial spin phases~\cite{zeiher_coherent_2017, samajdar_complex_2020, ebadi_quantum_2021} were realized and understood on a fundamental level. In recent years, the exploration of systems generating more complex patterns and phases became a leading research direction in many-body quantum physics. Particular focus hereby lies on quantum glasses~\cite{gopalakrishnan_frustration_2011, angelone_superglass_2016,hertkorn_pattern_2021, baio_multiple_2021}, many-body localization~\cite{smith_many-body_2016,choi_exploring_2016}, spin-liquids~\cite{zhou_quantum_2017,semeghini_probing_2021} and quasicrystals~\cite{viebahn_matter-wave_2019,mivehvar_emergent_2019,mendoza-coto_exploring_2021}. Realizing such complex systems establishes a path towards a deeper understanding of interacting quantum many-body systems, which ultimately fosters potential future applications in modern quantum technologies. Many of the effects mentioned above are related to frustration,~\ie the inability to satisfy all constraints imposed by the system's constituents at the same time. In this work we introduce and study a particularly clean and experimentally well controlled system which can give rise to frustration: a Rydberg-dressed BEC of neutral atoms trapped in an optical standing wave resonator, see~\fref{fig:setup}(a).  We elucidate some of the archetypal effects of the contest among the cavity-induced infinite range interactions and long-range Van-der-Waals interactions among Rydberg atoms and ultimately leading to glassy behavior. Previous work that combined Rydberg atoms and cavities focused on either generating large optical nonlinearities for single photons~\cite{grankin_quantum-optical_2015, boddeda_rydberg-induced_2016, gelhausen_quantum-optical_2016} or spin models on a lattice where the motional degrees of freedom are frozen~\cite{strack_dicke_2011, gopalakrishnan_frustration_2011, gelhausen_quantum-optical_2016}. Here, we focus on the spatial dynamics and resulting density self-ordering of the gas under competing long- and infinite-range interactions. Both constituents considered in this work exhibit a self-ordering phase transition due to roton mode softening~\cite{henkel_supersolid_2012, mottl_roton-type_2012}. These soft roton modes are a direct result of the respective long-range interactions. However, the two types of interactions have substantially different properties. The Rydberg dressing results in spherically symmetric long-range soft-core interactions [see~\fref{fig:setup}(b)] and the cavity photons induce anisotropic infinite-range periodic interactions between atoms [see~\fref{fig:setup}(c)]. This distinguishes our work from related studies focusing on a spherically symmetric interaction potential imposing two different length scales~\cite{barkan_controlled_2014, de_abreu_superstripes_2020, pupillo_quantum_2020}. In fact, while these works use a synthetic theoretically motivated long-range interaction potential, our study is based on an experimentally realistic configuration leading to the non-trivial physics presented below.

As we will show in the following, the particular properties of the light-matter coupled system,~\ie the competition between the two non-trivial interactions at different length scales and the nonlinear interaction between the cavity mode and the BEC lead to the dynamic formation of intriguing phases of quantum matter. In particular, we show how this system can be used to realize a stable glass phase,~\ie an amorphous solid with no long-range density order. The formation of this intriguing phase of quantum matter is attributed to geometrical frustration induced by the two competing interaction length scales and energies. Due to the additional superfluid properties of the BEC the resultant glassy phase is often also referred to as a superglass~\cite{boninsegni_superglass_2006,biroli_theory_2008, tam_superglass_2010,angelone_superglass_2016, angelone_nonequilibrium_2020}. The formation of this disordered glassy state in our system is not triggered by an externally imposed disorder or an external lattice geometry which results in frustration. It is the interplay and particular nature of the two considered model constituents (cavity photons in combination with Van-der-Waals interactions) which gives rise to the phases presented below.

\section{Model}\label{sec:model}
We consider a pancake-shaped BEC of $N$ atoms confined in two dimensions and trapped inside an optical standing-wave resonator, see~\fref{fig:setup}(a). The BEC atoms' ground state $\ket{g}$ is coupled via the cavity to a low-lying excited state $\ket{e}$ and, simultaneously, to a highly excited Rydberg $s$-state $\ket{r}$, realizing the V-configuration illustrated in~\fref{fig:setup}(a). The BEC atoms are pumped with a laser of frequency $\omega_p$ impinging from the side, transverse to the cavity axis, which drives the transition $\ket{g}\leftrightarrow\ket{e}$ at a Rabi frequency $\Omega_p$. This transition couples to a single cavity mode with a coupling strength $\mathcal{G}_p$. We consider the dispersive regime implying that the detuning $\Delta_p = \omega_p - \omega_{ge}$ is large compared to the Rabi frequency $\Omega_p$ ($\omega_{ge}$ is the transition frequency between $\ket{g}$ and $\ket{e}$). The transition $\ket{g}\leftrightarrow\ket{r}$ is driven by an additional laser (with frequency $\omega_R$) at a Rabi frequency $\Omega_R$ in the Rydberg dressing regime implying a large detuning $\Delta_R\gg\Omega_R$ with $\Delta_R\coloneqq \omega_R -\omega_{gr}$ ($\omega_{gr}$ is the transition frequency between $\ket{g}$ and $\ket{r}$). This allows the adiabatic elimination of the Rydberg state $\ket{r}$ resulting in a effective long-range two-body interaction potential~\cite{johnson_interactions_2010, honer_collective_2010, henkel_three-dimensional_2010, maucher_rydberg-induced_2011, balewski_rydberg_2014}
\begin{equation}
U_\mathrm{Ryd}(\rr,\rr') =  \frac{\tilde{C}_6}{R_c^6 + |\rr-\rr'|^6},
\label{eqn:dressing_pot}
\end{equation}
which arises due to the strong Van der Waals (VdW) interactions $\propto C_6/|\rr|^6$ ($\rr \in \mathbb{R}^2$) between Rydberg atoms. The parameters in~\eref{eqn:dressing_pot} are defined as $\tilde{C}_6\coloneqq \left(\frac{\Omega_R}{2\Delta_R}\right)^4 C_6$ and $R_c \coloneqq \left(-\frac{C_6}{2\hbar\Delta_R}\right)^{1/6}$. An exemplary plot of the functional dependence of this long-range interaction potential is shown in~\fref{fig:setup}(b). It was shown in previous work that a mean-field treatment suffices to capture the main features of cavity self-ordering~\cite{nagy_self-organization_2008, ritsch_cold_2013, mivehvar_cavity_2021} and Rydberg crystallization induced by the long-range interaction potential given in~\eref{eqn:dressing_pot}~\cite{henkel_three-dimensional_2010, henkel_supersolid_2012, hsueh_quantum_2012}. Therefore, we focus on a mean-field treatment. Here, we present the most important equations used throughout this work. A more detailed discussion of the model can be found in Appendix~\ref{app:mod_detail}. The dynamics is governed by two coupled equations. The equation for the BEC order parameter $\psi(\rr,t)$ is given as
\begin{subequations}
\begin{align}
&i\hbar\partial_t \psi(\rr,t) = \Big[-\frac{\hbar^2 \nabla^2}{2m} \nonumber\\
&+ \hbar U_0 |\alpha(t)|^2\cos^2(k_c x) + 2\hbar\eta \re[\alpha(t)]\cos(k_c x)\cos(k_c y) \nonumber\\
&+ \int_V d\rr' U_\mathrm{Ryd}(\rr,\rr')|\psi(\rr',t)|^2\Big]\psi(\rr,t),
\label{eqn:GPE_full}
\end{align}
and the dynamics of the mean-field cavity mode amplitude $\alpha(t)$ is governed by
\begin{equation}
i\partial_t\alpha(t) = \left[-\Delta_c + U_0 B[\psi] - i\kappa\right]\alpha(t) + \eta\theta[\psi].
\label{eqn:mode_dyn}
\end{equation}
\label{eqn:full_sys}
\end{subequations}
In~\eref{eqn:mode_dyn} we introduced the cavity bunching parameter $B[\psi]\coloneqq \int_V d\rr |\psi(\rr,t)|^2 \cos^2(k_c x)$ and the cavity mode order parameter $\theta[\psi] \coloneqq \int_V d\rr |\psi(\rr,t)|^2 \cos(k_c x)\cos(k_c y)$. The latter is the crucial quantity for understanding the self-ordering phase transition since the cavity mode can only take non-zero values if $\theta[\psi] \neq 0$, whereas the former only accounts for a cavity resonance frequency shift due to the BEC density. The potential depth of the cavity potential generated by inter-cavity photon scattering is defined as $U_0 \coloneqq \mathcal{G}_p^2/\Delta_p$, and the effective pump strength is $\eta\coloneqq \Omega_p\mathcal{G}_p/\Delta_p$ in~\eref{eqn:GPE_full}. In addition, $k_c = 2\pi/\lambda_c$ denotes the cavity wavenumber where $\lambda_c$ is the cavity resonance wavelength. We also introduced the detuning of the pump laser frequency $\omega_p$ with respect to the cavity resonance frequency $\omega_c$ as $\Delta_c\coloneqq \omega_p - \omega_c$ and the cavity decay rate $\kappa$.

\begin{figure}
\centering
\includegraphics[width=0.48\textwidth]{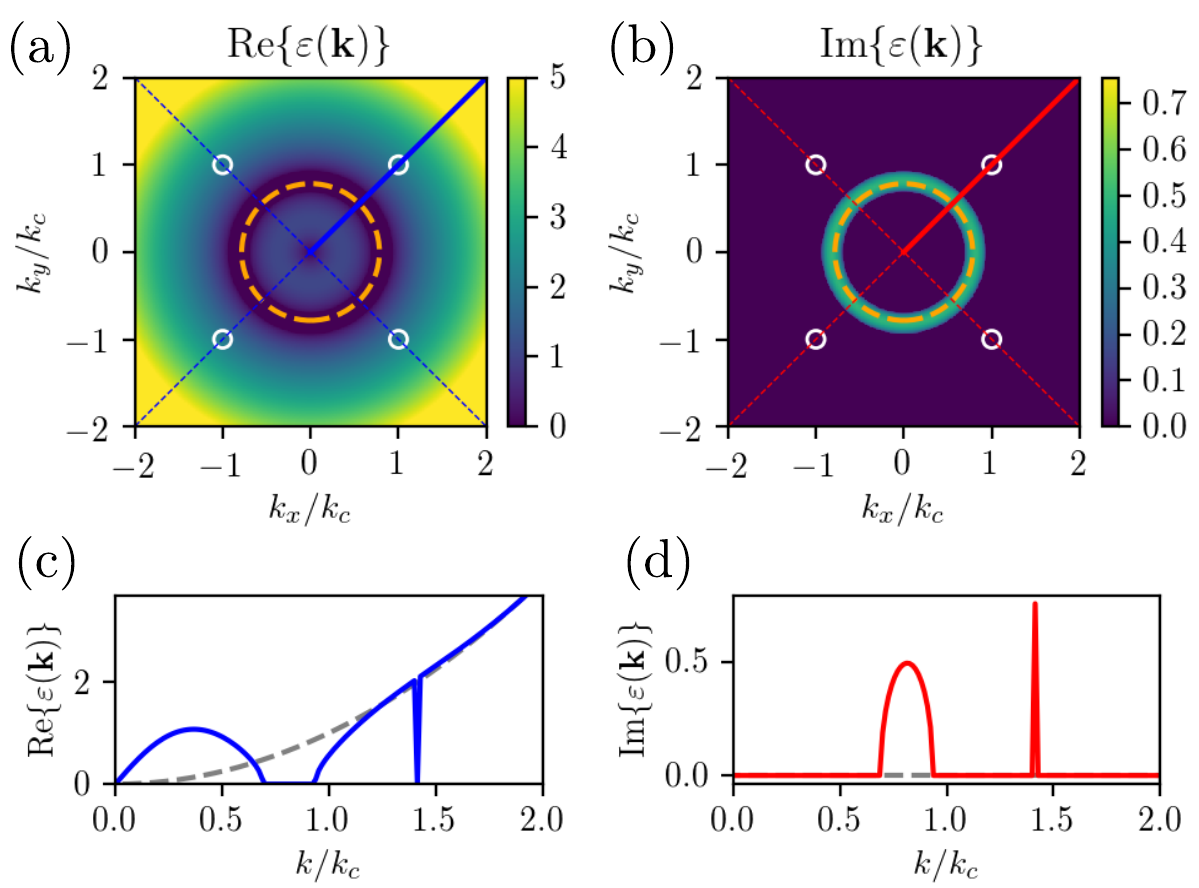} 
\caption{Real and imaginary part of the excitation spectrum given in~\eref{eqn:ex_spec} for $\tilde{C}_6 = 2.2 E_\mathrm{rec}$ and $\eta = 1.1\eta_\mathrm{crit}$,~\ie both interaction strengths above the threshold condition. The dashed orange circle in panels (a) and (b) marks the roton minimum induced by the spherically symmetric Rydberg interaction. The white circles indicate the four $\delta$-like rotons generated by the cavity interaction potential. Panels (c) and (d) show a cut along the diagonals for positive $k_x$ and $k_y$ values [red and blue lines in panels (a) and (b)]. The other parameters are: $R_c = 0.92 \lambda_c$, $\Delta_c = -10.0\omega_\mathrm{rec}$, $\kappa = 5.0\omega_\mathrm{rec}$ and $U_0 = -1.0\omega_\mathrm{rec}$.}
\label{fig:ex_spec_example}
\end{figure}

A particularly simple and insightful model can be obtained in the low-energy regime by adiabatically eliminating the cavity mode (for details,~\cf Appendix~\ref{app:mod_detail}), which results in a new effective equation for the BEC dynamics
\begin{align}
&i\hbar\partial_t \psi(\rr,t) = \Bigg[-\frac{\hbar^2}{2m}\nabla^2 \nonumber\\
&+\int_V d\rr'\left[U_\mathrm{cav}(\rr,\rr') + U_\mathrm{Ryd}(\rr,\rr')\right]|\psi(\rr',t)|^2\Bigg]\psi(\rr,t).
\label{eqn:elim_GPE}
\end{align}
In~\eref{eqn:elim_GPE} we introduced the cavity two-body interaction potential induced by the cavity photons as
\begin{equation}
U_\mathrm{cav}(\rr,\rr')= \hbar \mathcal{I}\cos(k_c x)\cos(k_c x')\cos(k_c y)\cos(k_c y'),
\label{eqn:cav_int_pot}
\end{equation}
where $\mathcal{I} \coloneqq \eta^2(\Delta_c - N U_0/2)/{[(\Delta_c - N U_0/2)^2 + \kappa^2]}$ is the effective interaction strength. An exemplary plot of this infinite-range interaction potential is shown in~\fref{fig:setup}(d). It is the competition between the two significantly different types of long-range interaction potentials $U_\mathrm{Ryd}$ and $U_\mathrm{cav}$ shown in~\ffref{fig:setup}(b) and (c), which contributes to the non-trivial emergent phases presented below. Note, that the simplified model given in~\eref{eqn:elim_GPE} provides a good intuitive picture about the fundamental interactions induced by the two model constituents. However, this approximate model does not cover all features arising from the nonlinear coupled dynamics of the cavity mode and the BEC [see~\eref{eqn:full_sys}], as discussed below.

\section{Roton instabilities}\label{sec:roton}
\begin{figure}
\centering
\includegraphics[width = 0.48\textwidth]{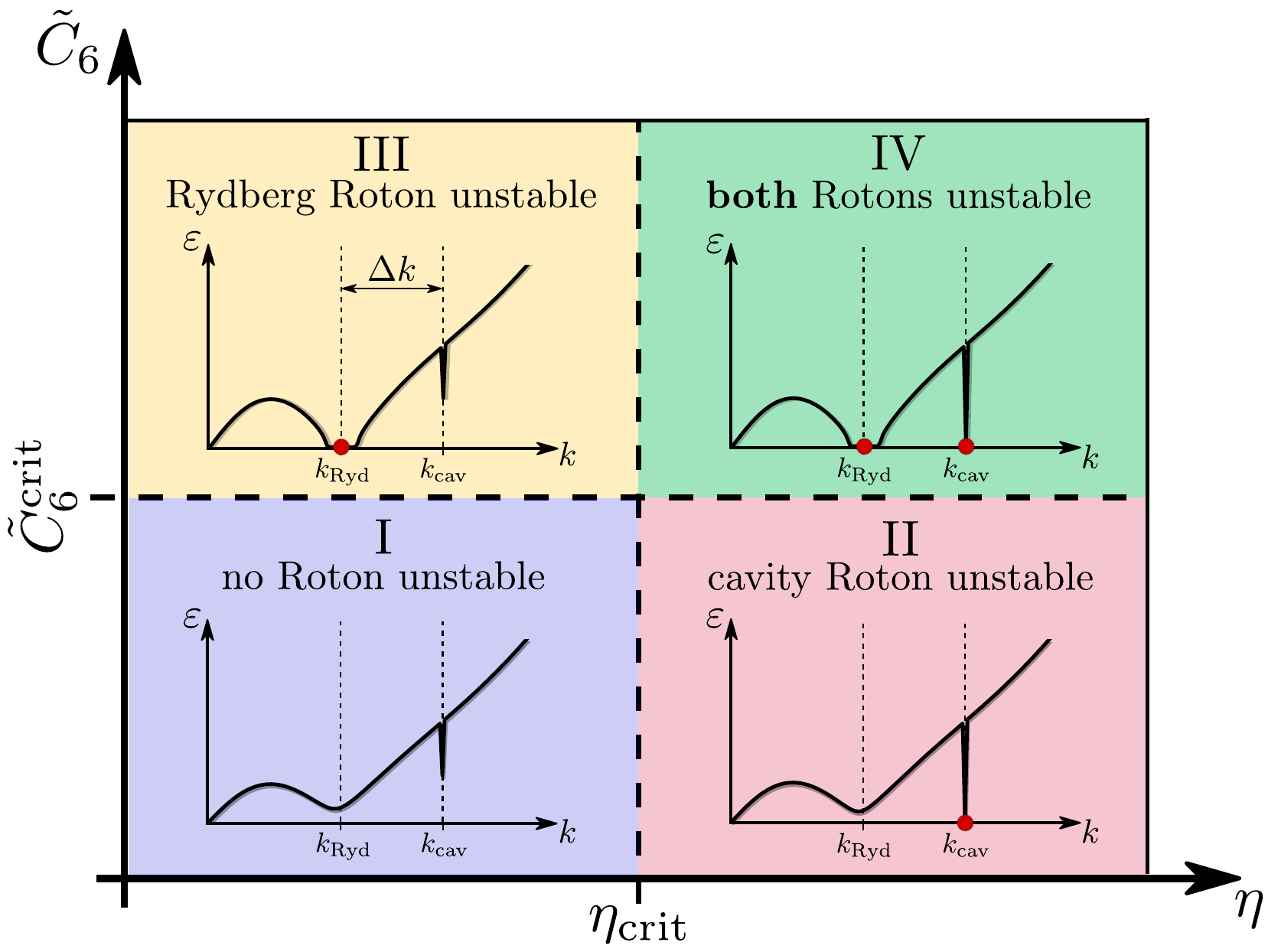} 
\caption{Sketch of the expected phase diagram based on the excitation spectrum. We distinguish four regions: I -- no roton softened,~\ie the homogeneous solution is stable, II -- the cavity roton touches zero (indicated by red circle),~\ie acquires an imaginary part, III --the Rydberg roton is unstable at $k=k_\mathrm{Ryd}$ and IV -- both rotons are unstable. The distance $\Delta k$ between the two $k$-values at which the rotons soften can be tuned by tuning $R_c$ in~\eref{eqn:dressing_pot}.}
\label{fig:PD_sketch}
\end{figure}
Roton-induced instabilities were first introduced and studied for superfluid Helium-4~\cite{landau_theory_1941} but they are a common feature of systems with long-range interactions. A dynamical instability of a certain nonzero $k$-mode in the collective excitation spectrum usually results in a phase transitions from homogeneous to periodic order. The Bogoliubov excitation spectrum for plane-wave excitations on top of a homogeneous condensate for a system with two-body long-range interaction potentials reads~\cite{pitaevskii_bose-einstein_2016}
\begin{equation}
\varepsilon (\kk) = \sqrt{\frac{\hbar^2 \kk^2}{2m}\left\{\frac{\hbar^2 \kk^2}{2 m} + 2\rho \left[U_\mathrm{Ryd}(\kk) + U_\mathrm{cav}(\kk)\right]\right\}},
\label{eqn:ex_spec}
\end{equation}
where $U_\mathrm{Ryd}(\kk)$ and $U_\mathrm{cav}(\kk)$ denote the Fourier Transforms (FTs) of the respective interaction potentials $U_\mathrm{Ryd}(\rr,\rr')$ and $U_\mathrm{cav}(\rr,\rr')$. We also introduced the particle density $\rho\coloneqq N/V$ with $V$ being the total volume of the BEC. The FT of the Rydberg interaction potential cannot be expressed in a comprehensive analytical formula. The FT of the cavity interaction potential~\eqref{eqn:cav_int_pot}, however, is given as $U_\mathrm{cav}(\kk) = \hbar\mathcal{I} \sum_{i,j \in\{0,1\}} \delta_{k_x,(-1)^i k_c}\delta_{k_y,(-1)^j k_c}$, where $\delta_{k_1,k_2}$ denotes the Kronecker-delta. In~\fref{fig:ex_spec_example} an exemplary excitation spectrum is shown. The spherically symmetric Rydberg interaction potential induces roton softening along a circle with a radius $k_\mathrm{Ryd} $ [indicated by the dashed orange line in~\fref{fig:ex_spec_example}(a) and (b)] . The cavity interaction potential, however, results in four $\delta$-shaped peaks/minima which are indicated by the white circles in~\fref{fig:ex_spec_example}(a) and (b), and are more prominently visible in panels~\fref{fig:ex_spec_example}(c) and (d). Hence, while the two interaction potentials considered in this work both result in dynamical instabilities induced by roton mode softening, the properties of the related excitation spectra differ significantly. The the cavity roton positions are always fixed by the cavity wavelength and are located at a value $k_\mathrm{cav}\coloneqq\sqrt{2}k_c$ while the value of $k_\mathrm{Ryd}$ can be tuned via the parameter $R_c$ in~\eref{eqn:dressing_pot}. Since each unstable $k$-mode can be associated with a characteristic length scale via $k_\mathrm{Ryd}=2\pi/l_\mathrm{Ryd}$ and $k_\mathrm{Ryd}=2\pi/l_\mathrm{cav}$, the role of different emergent length scales on the final steady or ground state can be analyzed by changing $R_c$ correspondingly. This can either be achieved by dressing to a different Rydberg state ($\tilde{C}_6$ scales like $n^{11}$, where $n$ is the principal quantum number) or by changing the detuning $\Delta_R$.

The homogeneous solution gets unstable toward a periodically ordered pattern if the excitation spectrum acquires an imaginary part at a non-zero $\kk$-value,~\ie $\min\left[\re(\varepsilon(\kk))|_{|\kk|>0}\right] = 0$. This results in two critical values $\tilde{C}_6^\mathrm{crit}$ (critical Rydberg interaction strength) and $\eta_\mathrm{crit}$ (critical effective cavity pump strength). Since the FT of the cavity interaction potential is solely given as the sum of four Kronecker-delta functions it is possible to find an analytical expression for the threshold for the pure cavity self-ordering case. Setting $\tilde{C}_6$ to zero and applying the threshold condition for $k = k_\mathrm{cav}$ in~\eref{eqn:ex_spec} results in
\begin{equation}
\eta_\mathrm{crit} = \sqrt{\frac{(\Delta_c - N U_0/2)^2 + \kappa^2}{N U_0 -2 \Delta_c}}2\sqrt{\omega_\mathrm{rec}},
\label{eqn:cav_SO_thres}
\end{equation}
which is the known threshold for self-organization of a BEC in a cavity in 2D~\cite{mivehvar_cavity_2021}.

\begin{figure}
\centering
\includegraphics[width = 0.48\textwidth]{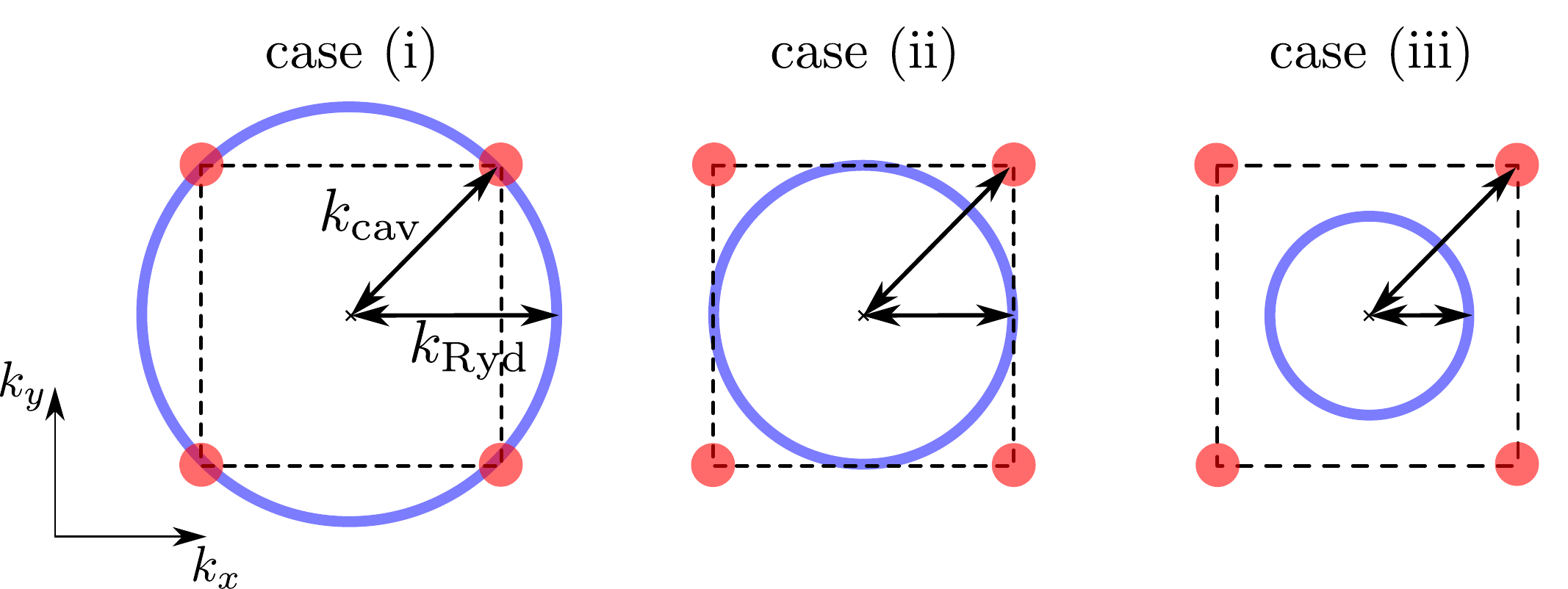} 
\caption{The three cases considered in this work. The red dots mark the four $k$-modes which are unstable due to the cavity interaction potential (see~\fref{fig:ex_spec_example}). The blue circle indicates all $k$-values which are unstable due to the Rydberg dressing potential. case (i): $k_\mathrm{Ryd} = k_\mathrm{cav}$, case (ii): $k_\mathrm{Ryd} = k_\mathrm{cav}/\sqrt{2}=k_c$ and case (iii): $k_\mathrm{Ryd} = k_\mathrm{cav}/2$. The cases (i) and (ii) represent special cases, whereas case (iii) corresponds to one example for the most general scenario.}
\label{fig:cases_sketch}
\end{figure}

These insights yield a first intuitive phase diagram shown in~\fref{fig:PD_sketch}. We expect the phase diagram to exhibit four different regions. In each region either no, one or two of the respective rotons are unstable (indicated by red circles in~\fref{fig:PD_sketch}). In addition to the two interaction strengths tuned in the phase diagram, the system features another relevant free parameter -- the Rydberg interaction range $R_c$. In the following we will restrict ourselves to three cases: (i) $k_\mathrm{Ryd} = k_\mathrm{cav}$ which is obtained by choosing $R_c = 0.51\lambda_c$, (ii) $k_\mathrm{Ryd} = k_\mathrm{cav}/\sqrt{2}=k_c$ with $R_c = 0.72\lambda_c$ and (iii) $k_\mathrm{Ryd} = k_\mathrm{cav}/2$ with $R_c = 1.2\lambda_c$. A sketch of the unstable roton length scales in $k$-space for all three cases is shown in~\fref{fig:cases_sketch}. We restrict ourselves to cases where $k_\mathrm{Ryd}<k_\mathrm{cav}$ because only in this case the additional Rydberg length scale competes with the length scale set by the cavity potential (see also Appendix~\ref{app:var_alg}). Note that the first two cases (i) and (ii) are special cases because the two wavenumbers are either the same [case (i)] or $k_\mathrm{Ryd}$ is commensurate with $k_c$, \ie the fundamental length scale along the cavity axis [case (ii)]. Case (iii), however, corresponds to one example of the most general incommensurate scenario. We chose a factor two for the wavenumber ratio in case (iii) but the same qualitative results as presented below were obtained for a variety of different wavenumber ratios.

\section{Self-consistent ground state}\label{sec:sc_GS}
One major element of the studied system is the interplay between the dynamic cavity field and the BEC dynamics. While the simplified model given in~\eref{eqn:elim_GPE} provides a good intuitive picture about the features and roles of the two different interactions, it does not fully capture the competing time scales in the system. Hence, we now analyze the full system given in~\eref{eqn:full_sys}. We calculate the self-consistent ground state for the set of~\eeref{eqn:full_sys} via a variational approach. To this end, we employ a self-consistent iterative algorithm: 1) choose a random initial steady-state value $\alpha_{ss}$. 2) plug this value into the GPE equation~\eqref{eqn:GPE_full} and perform an imaginary time evolution ($t\rightarrow -i\tau$) to find the lowest energy state. 3) calculate a new steady-state of~\eref{eqn:mode_dyn} by setting $\partial_t \alpha(t) = 0$ and solving for $\alpha$. We iterate until we find a state where $|\alpha_{ss}|$ lies within a convergence radius of $10^{-6}$. Note that this self-consistent algorithm is not necessarily convex. In general, the energy functional
\begin{align}
\mathcal{E}[\psi] &= \int_V d\rr \Bigg[-\frac{\hbar^2\nabla^2}{2m} + \hbar U_0 |\alpha(t)|^2\cos^2(k_c x)\nonumber\\
&+ 2\hbar\eta \re[\alpha(t)]\cos(k_c x)\cos(k_c y) \nonumber\\
&+ \int_V d\rr' U_\mathrm{Ryd}(\rr,\rr')|\psi(\rr')|^2 \Bigg]|\psi(\rr)|^2,
\label{eqn:E-functional}
\end{align}
which results in the GPE in~\eref{eqn:GPE_full}, can have multiple local minima for a given value of $\alpha$. Whether one finds the global minimum or not depends on the initial condition for the imaginary time evolution. Therefore, we perform this algorithm for various different initial conditions and compare the obtained values of $|\alpha|$ and the corresponding ground state energies by evaluating the energy functional in~\eref{eqn:E-functional}. Only if all initial conditions result in the same energy and mode amplitude the energy minimization problem is convex. For details we refer to Appendix~\ref{app:var_alg}.

\subsection{Case (i) --- equal length scales}
\begin{figure}
\centering
\includegraphics[width=0.48\textwidth]{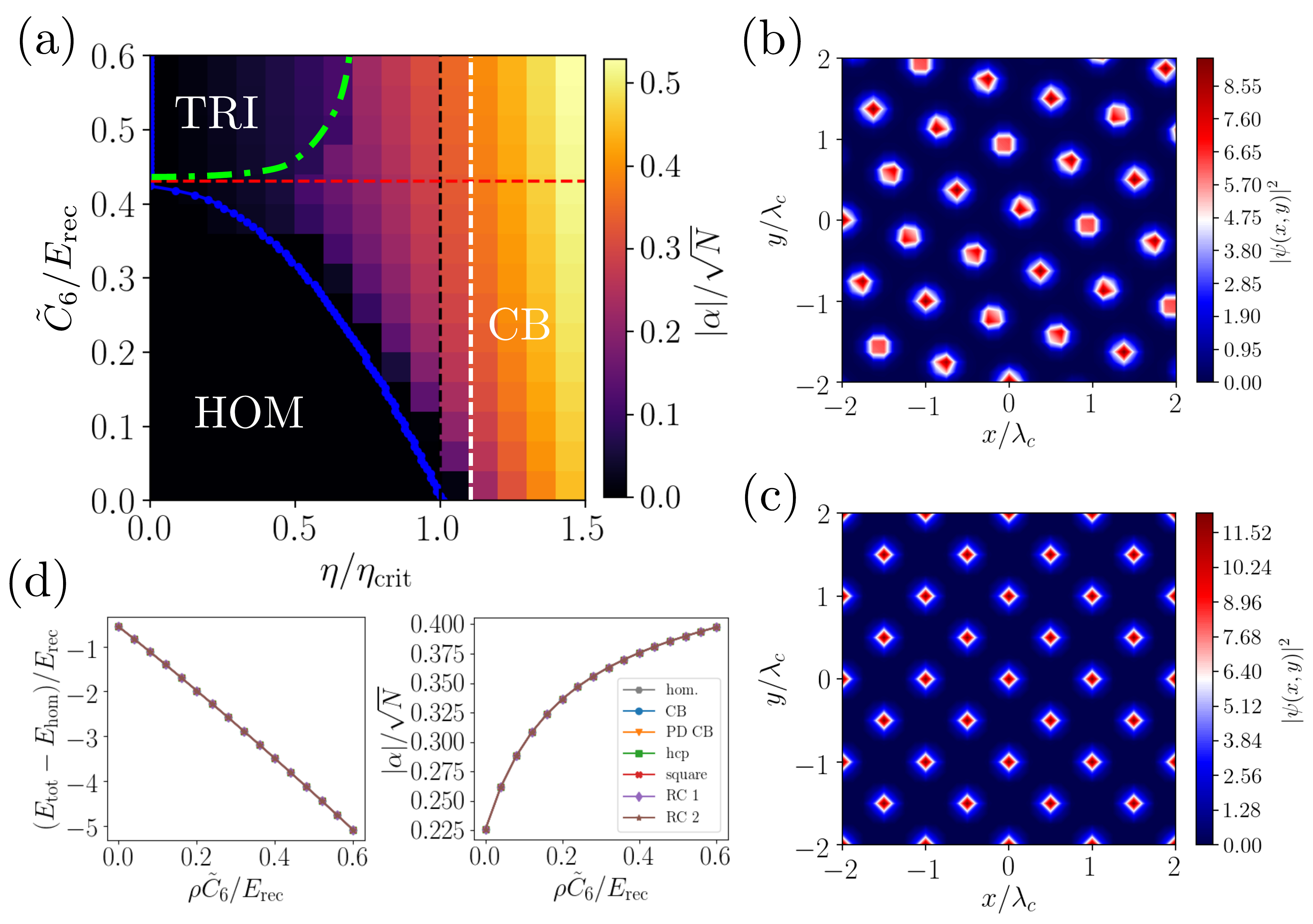}
\caption{(a) Self-consistent ground state phase diagram for case (i) --- $k_\mathrm{Ryd}$ = $k_\mathrm{cav}$ ($R_c =0.51\lambda_c$). The dashed black and red curves indicate the critical values for independent Ryd\-berg crystallization (horizontal red dashed) or cavity self-organization (vertical black dashed). The blue curve is the critical value at which the excitation spectrum in~\eref{eqn:ex_spec} touches zero for the given parameters. The green dash-dotted curve is a guide to the eye indicating the phase boundary. (b) $|\alpha|$ values and ground-state energies (shifted by the energy of the homogeneous solution $E_\mathrm{hom}$) for values along the white dashed line in panel (a) for all seven  initial conditions. (c) Exemplary density distribution for the triangular (TRI) lattice phase with $\eta = 0.2 \eta_\mathrm{crit}$ and $\tilde{C}_6 = 0.6 E_\mathrm{rec}$. (d) Exemplary density distribution for the checkerboard (CB) lattice phase with $\eta = 1.1 \eta_\mathrm{crit}$ and $\tilde{C}_6 = 0.6 E_\mathrm{rec}$. All other parameters as in~\fref{fig:ex_spec_example}.}
\label{fig:PD_checkerboard}
\end{figure}
\begin{figure*}
\centering
\includegraphics[width=1.0\textwidth]{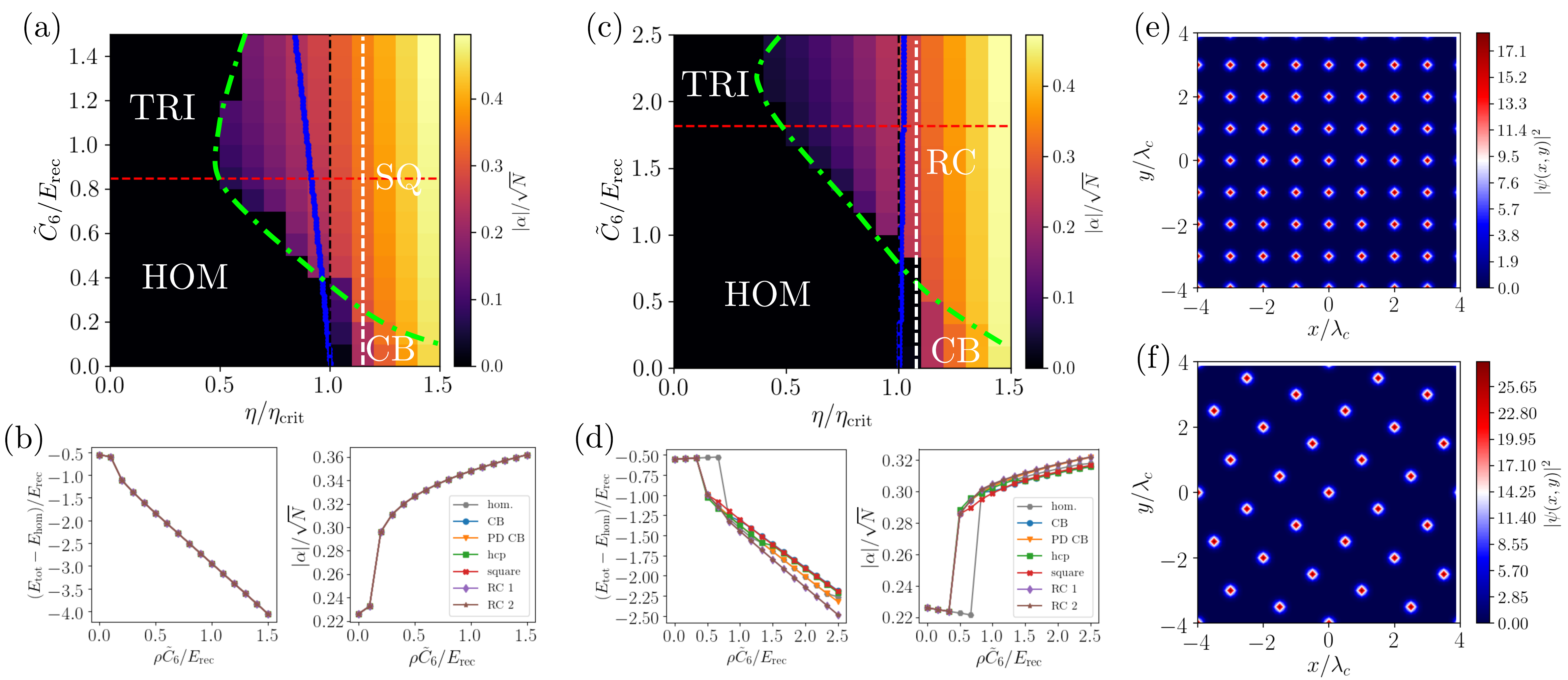} 
\caption{(a) Numerical phase diagram for case (ii) --- $k_\mathrm{Ryd}$ = $k_\mathrm{cav}/\sqrt{2}$ ($R_c =0.72\lambda_c$). (b) Ground-state energy and cavity mode amplitude $|\alpha|$ along the white vertical dashed line in panel (a) for all seven initial conditions. (c) Numerical phase diagram for case (iii) --- $k_\mathrm{Ryd}$ = $k_\mathrm{cav}/2$ ($R_c = 1.2\lambda_c$). (d) Same as (c) but for case (iii). The different initial conditions no longer result in the same ground-state energies and mode amplitudes. The iterative energy minimization is not convex. (e) Exemplary ground-state density distribution for the square (SQ) phase for case (ii). (f) Exemplary ground state density distribution for the rotated chain phase for case (iii). Note that the same density distribution rotated by $\pi/2$ is a ground state as well, as the degeneracy between RC 1 and RC 2 in panel (d) indicates. The lines on top of the color coding in panels (a) and (c) and all other parameters are the same quantities as in~\fref{fig:PD_checkerboard}(a).}
\label{fig:PD_square_glass}
\end{figure*}
We first consider the special case where $k_\mathrm{Ryd}=k_\mathrm{cav}$ (see~\fref{fig:cases_sketch}), which implies equal length scales $l_\mathrm{Ryd} = l_\mathrm{cav}$ ($R_c = 0.51\lambda_c)$. In this case, the regions II and IV in the phenomenological phase diagram in~\fref{fig:PD_sketch} are expected to merge. This is verified by the numerical phase diagram shown in~\fref{fig:PD_checkerboard}(a). It exhibits three different phases a homogeneous phase (HOM), a triangular phase (TRI) [see~\fref{fig:PD_checkerboard}(b)] and a checkerboard lattice phase (CB) [see~\fref{fig:PD_checkerboard}(c)]. The TRI state only forms if the Rydberg induced roton is unstable,~\ie beyond the critical Rydberg interaction strength indicated by the red dashed line in~\fref{fig:PD_checkerboard}(a). For increasing cavity pump, the system organizes in a checkerboard pattern, which is the configuration realized in cavity self-organization without Rydberg dressing~\cite{ritsch_cold_2013, mivehvar_cavity_2021}. Note, however, that the softening of the Rydberg roton affects the threshold for cavity self-ordering (blue line in~\fref{fig:PD_checkerboard}(a)). This implies that the effect of the Rydberg roton can be observed already far below the critical Rydberg interaction strength $\tilde{C}_6^\mathrm{crit}$. This is a remarkable result as the observation of density modulation due to Rydberg dressing has proven elusive in experiments~\cite{balewski_rydberg_2014} due to experimental complications such as unfavorable scaling with the density and long time scales required for experiments. Our results suggest a way to observe effects of Rydberg dressing indirectly for much smaller Rydberg fractions (\ie smaller $\tilde{C}_6$). In~\fref{fig:PD_checkerboard}(d) we show the obtained ground state energy obtained from~\eref{eqn:E-functional} and the cavity mode amplitude for all seven employed initial conditions. We shifted all energies with respect to the homogeneous state energy $E_\mathrm{hom}$ which is obtained by evaluating~\eref{eqn:E-functional} for $\alpha = 0.0$ and a homogeneous BEC density. Energies are given in units of the recoil energy $E_\mathrm{rec}\coloneqq \hbar^2k_c^2/(2m) = \hbar \omega_\mathrm{rec}$ throughout this work. We find that all curves for all initial conditions lie on top of each other, implying that this particular iterative energy minimization for case (i) is convex in $\psi$ and $\alpha$.

\subsection{Cases (ii) and (iii) --- different length scales}
The main property of the considered system is that two interactions with different functional dependence \emph{and} different intrinsic length scale are combined. Here, we show that this feature results in highly non-trivial physics which goes beyond what is achievable with only a single model constituent.~\fref{fig:PD_square_glass} exhibits the two obtained ground state phase diagrams together with the additional non-trivial density distributions realized in these regimes. Since now two fundamental length scales become unstable the phase diagram resembles the intuitive picture from~\fref{fig:PD_sketch} and contains four different phase regions. The critical values obtained via~\eref{eqn:ex_spec} in section~\ref{sec:roton}, however, do not coincide with the numerical values. This shows that the simplified model discussed in section~\ref{sec:roton} for the adiabatically eliminated cavity dynamics [see~\eref{eqn:elim_GPE}] provides a good first intuition about the contributing interactions and the expected phase diagram, but fails for different competing length scales due to a break down of the adiabatic solution. This is due to the influence of the additional Rydberg length scale on the mode dynamics which is taken into account in the full numerics resulting in the phase diagrams shown in~\fref{fig:PD_square_glass}. The instability in this system is actually induced by the combined collective instability of the BEC density \emph{and} the cavity mode amplitude. The fluctuations of the BEC couple to the fluctuations of the cavity mode and vice versa. This results in actual thresholds for the transitions $\mathrm{TRI} \leftrightarrow \mathrm{SQ}$ and $\mathrm{CB} \leftrightarrow \mathrm{SQ}$ in~\fref{fig:PD_square_glass}(a) and $\mathrm{TRI} \leftrightarrow \mathrm{RC}$ and $\mathrm{CB} \leftrightarrow \mathrm{RC}$ in~\fref{fig:PD_square_glass}(c) that are below the values anticipated by the simplified excitation spectrum in~\eref{eqn:ex_spec}. This again could facilitate the experimental observation of effects induced due to Rydberg induced roton mode softening at smaller Rydberg interaction strengths $\tilde{C}_6$.

For case (ii) we find that, in addition to the TRI and CB phases, a square (SQ) lattice phase emerges [\cf~\fref{fig:PD_square_glass}(e)]. This phase is a direct result of the non-trivial interplay between the two different interactions at these particular length scales. All seven initial conditions again result in the same ground state energies and mode amplitudes [see \fref{fig:PD_square_glass}(b)]. This feature changes in the incommensurate case (iii) --- $k_\mathrm{Ryd} = k_\mathrm{cav}/2$. We identify an additional rotated chain (RC) phase [see~\fref{fig:PD_square_glass}(c) and (f)], where the iterative energy minimization in the self-consistent algorithm is no longer convex. There are two possible realizations of the RC phase. In~\fref{fig:PD_square_glass}(f), we show one possible realization. However, the given density pattern rotated by $\pi/2$ is another ground state with the same energy. This can be seen by the two degenerate lowest energy curves in~\fref{fig:PD_square_glass}(d) which correspond to the two possible realizations of the RC phase. The two degenerate states maximize the population of the cavity mode [see~\fref{fig:PD_square_glass}(d)] while fulfilling the length scale restrictions imposed by the Rydberg interactions,~\cf Appendix~\ref{app:var_alg} for details. Such degeneracy in combination with non-convex features is a prime indicator for geometrical frustration ultimately leading to glassiness as we show in the next section. Note that we restricted our analysis on three cases for a concise presentation of the results. It should be mentioned that cases (i) and (ii) are indeed two special cases, which lead to a convex problem. Any other choice (case (iii) is one possibility) of length scale ratios results in a non-convex energy landscape. The non-convex nature of the energy landscape in case (iii) is the crucial feature which results in the superglass formation presented below.

\section{Dynamics}\label{sec:dynamics}
It is a priori unclear whether the self-consistent ground state can be reached by dynamically evolving ~\eref{eqn:full_sys} in real time. While this is usually the case for systems with a unique lowest-energy state, ground-state degeneracy in combination with higher lying energy states may block the dynamical realization of the RC phase in case (iii). In fact, we will show below that the steady-state for case (iii) is an amorphous solid with no long-range density order.

\begin{figure}
\centering
\includegraphics[width = 0.49\textwidth]{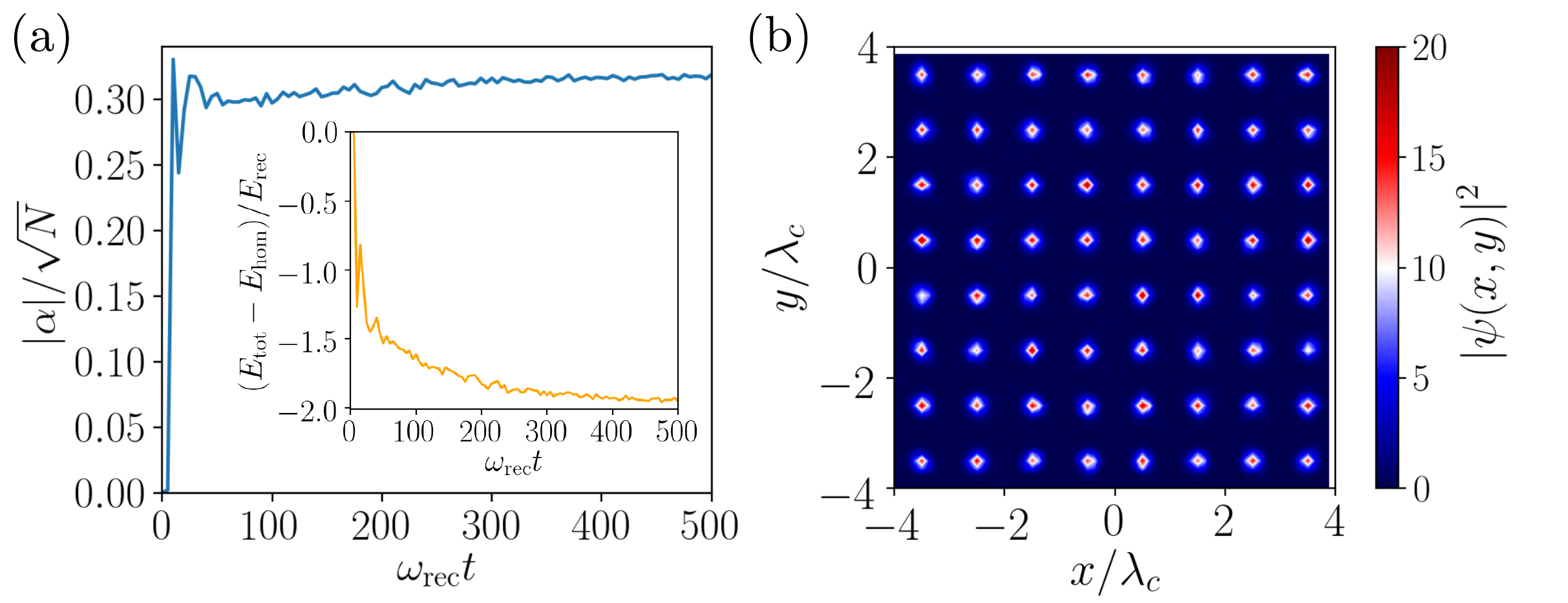} 
\caption{(a) Time evolution of the cavity mode amplitude for case (ii). The inset shows the corresponding time evolution of the total energy obtained via~\eref{eqn:E-functional}. (b) Steady-state at $t = 500/\omega_\mathrm{rec}$. Parameters: $\eta = 1.1\eta_\mathrm{crit}$, $\tilde{C}_6 = 1.0 E_\mathrm{rec}$, $R_c = 0.72\lambda_c$. All other parameters as in~\fref{fig:ex_spec_example}.}
\label{fig:square_ss}
\end{figure}

In~\fref{fig:square_ss}, the time evolution of the cavity mode amplitude and the total energy together with the steady-state density distribution at $t = 500/\omega_\mathrm{rec}$ is shown for case (ii). We find that the square lattice obtained in the previous section is a stable steady state, which can be obtained dynamically when starting from a homogeneous BEC.
\begin{figure}
\centering
\includegraphics[width = 0.48\textwidth]{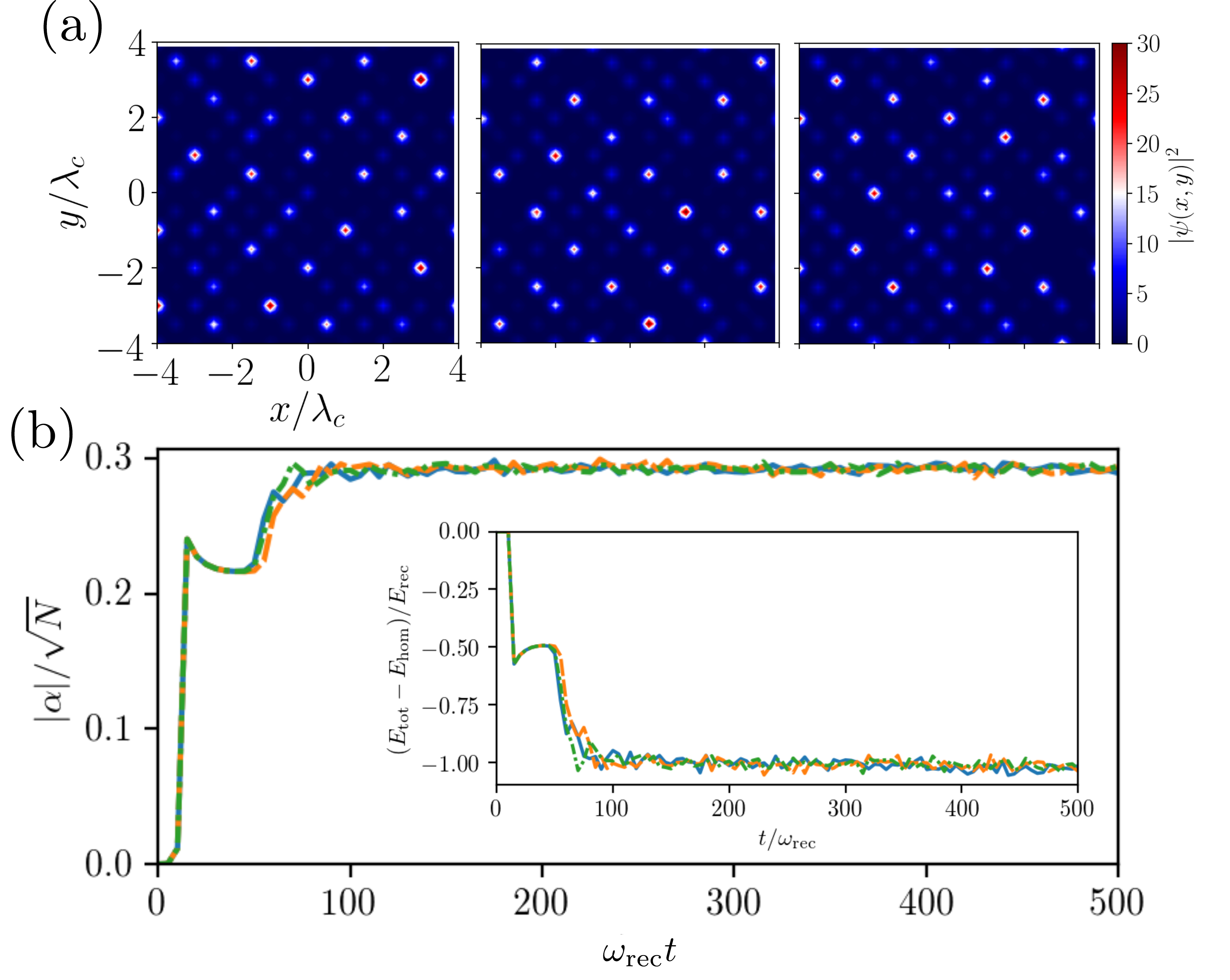} 
\caption{(a) Three different steady-states for three runs for case (iii). Each run generates a different density pattern with no long range order. This is a typical indication of an amorphous solid (or a glass phase). All realized patterns, however, scatter exactly the same amount of light into the cavity as it can be seen from panel (b).  Parameters: $\eta = 1.1\eta_\mathrm{crit}$, $\tilde{C}_6 = 1.5 E_\mathrm{rec}$, $R_c = 1.2\lambda_c$. All other parameters as in~\fref{fig:ex_spec_example}.}
\label{fig:glassy_ss}
\end{figure}
In contrast, the steady state for case (iii) is stable but \emph{not} the rotated chain state as the self-consistent phase diagram suggests. The steady state is an inhomogeneous density pattern with no long-range density order. In~\fref{fig:glassy_ss}(a), we show three outcomes for three independent runs of the time evolution for the same parameters. Every run results in a different density distribution. This feature, in addition to the BEC's superfluidity, qualifies this phase as a so-called superglass~\cite{boninsegni_superglass_2006,biroli_theory_2008, tam_superglass_2010}. The glassy behavior can be directly understood from the nonconvex properties and the degeneracy of the energies found in the previous section [see~\fref{fig:PD_square_glass}(f)]. Each run results in a different mixture of the degenerate ground states giving rise to the glassy steady state. Remarkably, however, the resulting  different density distributions still have exactly the same steady-state cavity mode amplitude. Hence, the different glassy states are not distinguishable via the cavity fields. Note that the formation of this superglass state is a direct consequence of the two particularly chosen model constituents and cannot be straightforwardly realized by combining \emph{any} two long-range interaction potentials.

\section{Conclusions and Outlook}\label{sec:conclusion}

We showed that combining the spherically symmetric long-range interactions induced in a Rydberg-dressed BEC with infinite-range cavity-induced interactions establishes a versatile platform for studying the interplay between different interaction types and landscapes on the phases of quantum matter. Most strikingly, we point out a way towards realizing a superglass phase in a robust and well controlled system. The disorder in this superglass phase is not imprinted from outside via some external potential or lattice but it is solely driven by quantum fluctuations and the interplay of the two interaction potentials. This dynamical, fluctuation induced formation of a superglass phase is closely related to the theoretically predicted phase transition from a liquid to a superglass after a temperature quench in Helium-4~\cite{boninsegni_superglass_2006}, which has not been realized experimentally or found elsewhere yet. Our findings hint towards a realizable system to dynamically study phase transitions from a homogeneous or solid (CB phase) to a superglass by tuning the Rydberg interaction strength. The studied system is highly non-linear and the formation of the intriguing phases presented in this work involves the interplay between the BEC wave function $\psi$ \emph{and} the cavity mode $\alpha$. This gives way to ease the observation of Rydberg induced instabilities experimentally because the non-linear coupling of atomic density fluctuations to the cavity mode facilitates the softening of the Rydberg roton. Our findings also have potential applications in modern quantum technologies: The formation of a density ordered pattern in this hybrid system ultimately resembles a quantum optimization problem. Hence, the system could be tailored to particular optimization problems. In addition, glassy states are promising candidates for efficient quantum memories.

Our work lays the ground for a variety of further studies combining long-range interacting systems theoretically as well as experimentally. This research avenue is fostered by the new experimental and theoretical opportunities opening up in recently established experimental setups all over the world. Our proposed setup should be realizable in state-of-the-art cavity QED setups. In particular, the realization of the glassy phase where the parameters are favorable for experiments could readily be realized. However, the setup presented here is not the only system which is expected to exhibit such intriguing properties. Other promising avenues are combining other systems exhibiting light-induced instabilities in free space~\cite{ostermann_spontaneous_2016, firth_thick-medium_2017, ostermann_probing_2017, dimitrova_observation_2017, zhang_long-range_2018, zhang_self-bound_2021, baio_multiple_2021} with Rydberg dressing or long-range dipole interactions. More control over a wide range of realizable density patterns is expected if one assumes even more complex cavity-induced interaction potentials or Rydberg interactions. This can be achieved by changing to more complex cavity geometries such as multi-mode resonators~\cite{keller_quenches_2018,guo_sign-changing_2019, guo_emergent_2019} or dressing to spherically asymmetric Rydberg $p$-states~\cite{glaetzle_quantum_2014,glaetzle_designing_2015}. From a quantum optical viewpoint the dynamic transverse coupling of Rydberg atoms to a cavity could result in a highly nonlinear mechanism to enhance the achievable nonlinearities with state-of-the-art setups. The platform presented in this work could also be extended to include spin degrees of freedom, and, serve as a viable tool to gain deeper understanding of spin glasses or even spin liquids~\cite{sherrington_solvable_1975,semeghini_probing_2021, kong_melting_2021}.

While this work focused on highlighting the remarkable features of this system and their intriguing consequences, several directions remain open for further research. One open and exciting question is the connection of the formation of the glassy phase to Anderson localization or even many-body localization. Also, the role of beyond mean-field effects and temperature needs further investigation. To answer these questions, alternative theoretical techniques, which go beyond the scope of the present manuscript, have to be developed and applied~\cite{boninsegni_worm_2006, boninsegni_worm_2006_2, lin_superfluidmott-insulator_2019, rui_mott_2021}. In any case, the results presented in this work open up exciting avenues in the growing research field of hybrid quantum systems with long-range interactions.

\acknowledgments{
We would like to thank Helmut Ritsch for fruitful discussions. S.F.Y. would like to acknowledge funding by the AFOSR (applications of dense atomic media), the NSF (theory of dense atomic media), and the DOE (structure of 2D atomic media). S.O. is supported by a postdoctoral fellowship of the Max Planck - Harvard Research Center for Quantum Optics (MPHQ). V.W. acknowledges support from the NSF through a grant for the Institute for Theoretical Atomic, Molecular, and Optical Physics at Harvard University and the Smithsonian Astrophysical Observatory.

The numerical simulations were performed with the open-source framework \texttt{QuantumOptics.jl}~\cite{kramer_quantumopticsjl_2018}.}

\appendix

\section{Detailed model derivation}\label{app:mod_detail}
\begin{figure*}
    \centering
    \includegraphics[width = 0.99\textwidth]{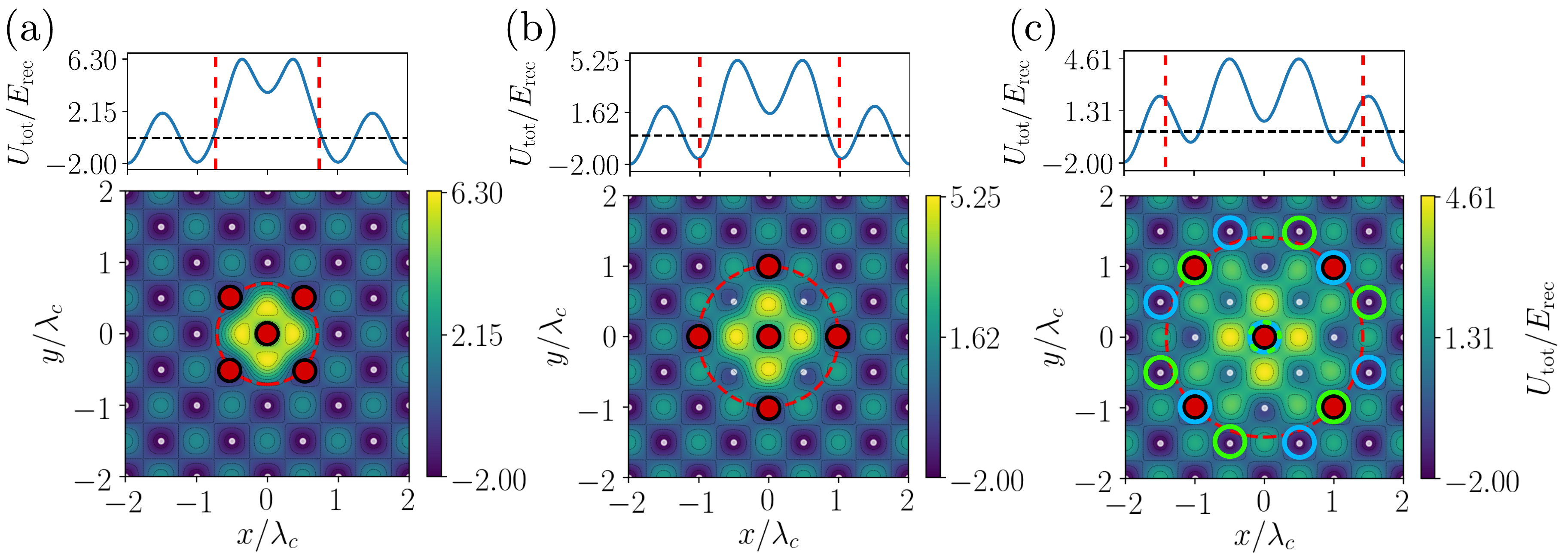}
    \caption{Exemplary total interaction potentials from~\eref{eqn:comb_int_pot} for the three cases treated throughout this work. (a) Case (i); $\eta = 1.0\eta_\mathrm{crit}$ and $\tilde{C}_6 = 0.1 E_\mathrm{rec}$, (b) Case (ii); $\eta = 1.0\eta_\mathrm{crit}$ and $\tilde{C}_6 = 0.5 E_\mathrm{rec}$), (c) Case (iii); $\eta = 1.0\eta_\mathrm{crit}$ and $\tilde{C}_6 = 3.0 E_\mathrm{rec}$. The red dashed lines indicate the length scale imposed by the Rydberg roton. In panels (a) and (b) the red dots mark the first potential minima fulfilling the length scale restrictions imposed via the soft-core interaction potential. In panel (c) the light green and blue circles mark additional configurations fulfilling this restriction. This gives rise to a larger number of potential self-consistent ground states ultimately leading to the glassy behavior discussed in the main text. The dashed red lines indicate the length scale imposed by the Rydberg interaction potential for the respective cases.}
    \label{fig:comb_pot}
\end{figure*}

In the following we provide a concise derivation of the model presented in section~\ref{sec:model}. For a more detailed discussion of the underlying physics we refer to Refs.~\cite{maschler_ultracold_2008} and~\cite{henkel_three-dimensional_2010}. The full many-body Hamiltonian for the considered system can be written as the sum of five Hamiltonians $H = H_a + H_{a-c} +  H_c + H_\mathrm{int}^\mathrm{2-b.} + H_\mathrm{int}^\mathrm{Ryd}$ with
\begin{subequations}
\begin{align}
H_a &= \int d\rr \Psi^\dagger(\rr)\left[-\frac{\hbar^2\nabla^2}{2m} + \hbar U_p \cos^2(k_cy)\right]\Psi(\rr),\label{eqn:MB_at}\\
H_{a-c} &= \int d\rr \Psi^\dagger(\rr)\big[\hbar U_0 \cos^2(k_cx)a^\dagger a \nonumber \\
&+ \hbar \eta \cos(k_c x)\cos(k_c y)(a+a^\dagger) \big] \Psi(\mathbf{r}), \label{eqn:MB_a-c}\\
H_c &= -\hbar \Delta_c a^\dagger a, \label{eqn:MB_c}\\
H_\mathrm{int}^\mathrm{2-b.} &= \frac{g}{2} \int d\rr \Psi^\dagger(\rr)\Psi^\dagger(\rr)\Psi(\rr)\Psi(\rr) \label{eqn:MB_s-int}\\
H_\mathrm{int}^\mathrm{Ryd} &= \iint d\rr d\rr'\Psi^\dagger(\rr)\Psi^\dagger(\rr')\frac{\tilde{C}_6}{R_c^6+|\rr- \rr'|^6}\Psi(\rr')\Psi(\rr).\label{eqn:MB_Ryd-int}
\end{align}
\label{eqn:MB_H}
\end{subequations}
Here, $\Psi$ ($\Psi^\dagger$) are annihilation (creation) operators of bosonic ground state atoms,~\ie $\left[\Psi(\rr),\Psi^\dagger(\rr')\right] = \delta(\rr-\rr')$ and $a$ ($a^\dagger$) are annihilation (creation) operators of the photonic cavity mode fulfilling $\left[a,a^\dagger\right]= 1$.  $U_p\coloneqq \Omega_p^2/\Delta_p$ is the potential depth of the lattice generated by the two interfering pump beams and $U_0 \coloneqq \mathcal{G}_p^2/\Delta_p$ is the depth of the optical potential generated by photon scattering from the atomic density distribution inside the cavity. The effective cavity pump strength is given as $\eta\coloneqq \Omega_p\mathcal{G}_p/\Delta_p$ and $k_c = 2\pi/\lambda_c$ denotes the wavenumber of the cavity mode. The Hamiltonian in~\eref{eqn:MB_s-int} takes into account local two-body interactions between atoms and is omitted throughout this work. This is a reasonable assumption since this interaction strength can be tuned to be small via,~\eg a Feshbach resonance such that cavity and Rydberg induced long-range interactions dominate the dynamics~\cite{brennecke_real-time_2013, henkel_three-dimensional_2010}. The dynamics of the hybrid atom-cavity system is governed by the Heisenberg-Langevin equations $i\hbar\partial_t \Psi(\rr,t) = \left[\Psi(\rr,t),H\right]$ and $i\hbar\partial_t a(t) = \left[a(t),H\right] - i\hbar\kappa a(t)$ where the decay of the cavity mode at a rate $\kappa$ is included. Calculating these equations of motion and performing a mean-field approximation via $\Psi(\rr,t)\rightarrow \langle\Psi(\rr,t)\rangle \coloneqq \psi(\rr,t)$ and $a(t)\rightarrow\langle a(t)\rangle \coloneqq \alpha(t)$, results in the two coupled c-number equations for the BEC order parameter $\psi(\rr,t)$ and the cavity mode amplitude $\alpha(t)$ given in~\eref{eqn:full_sys} of the main text.

To adiabatically eliminate the cavity mode we calculate the equation of motion for the field operator $a$ and solve for its steady-state ($\partial_t a = 0$). This results in
\begin{equation}
a_\mathrm{ss} = \frac{\eta \int d\rr \Psi^\dagger(\rr,t) \cos(k_c x)\cos(k_c y)\Psi(\rr,t)}{\left[\Delta_c - U_0 \int d\rr \Psi^\dagger(\rr,t) \cos^2(k_c x) \Psi(\rr,t)\right] + i\kappa}.
\end{equation}
Apart from the fundamental cavity parameters $\eta$, $\Delta_c$ and $\kappa$ the steady-state value for the cavity mode is determined by the two quantities
\begin{align}
\Theta[\Psi(\rr,t)] &\coloneqq \int d\rr \Psi^\dagger(\rr,t) \cos(k_c x)\cos(k_c y)\Psi(\rr,t),\\
\mathcal{B}[\Psi(\rr,t)] &\coloneqq \int d\rr \Psi^\dagger(\rr,t) \cos^2(k_c x) \Psi(\rr,t).
\end{align}
These two quantities obviously depend on the BEC state which exhibits the non-trivial coupling between the cavity mode and the BEC. While $\mathcal{B}$ only acts as an effective shift of the cavity resonance frequency that comes into play as soon as $a_\mathrm{ss}\neq 0$, $\Theta$ is the crucial parameter when it comes to understanding the cavity self-ordering phase transition. The cavity mode is nonzero if $\Theta \neq 0$. Hence this parameter is crucial for the instability described in the main text. To simplify the model we replace $\mathcal{B}$  with its value for the homogeneous condensate $\mathcal{B} = N/2$ while keeping the full functional dependence of $\Theta$. To eliminate the cavity field we plug the resultant steady-state solution into the many-body Hamiltonian. Since we are solely interested in a model capturing the dynamic instability which is governed by terms $\propto \cos(k_c x)\cos(k_c y)$ (see argument above) we keep only these terms. This results in the effective interaction Hamiltonian
\begin{widetext}
    \begin{equation}
    H^{\mathrm{cav}}_\mathrm{int} = \hbar \mathcal{I} \iint d\rr d\rr' \Psi^\dagger(\rr)\Psi^\dagger(\rr')\cos(k_c x)\cos(k_c x')\cos(k_c y)\cos(k_c y')\Psi(\rr')\Psi(\rr).
    \label{eqn:Heff}
    \end{equation}
\end{widetext}
where we used the symmetry of the cosine function in the second line and introduced the cavity induced interaction strength as
\begin{equation}
\mathcal{I} = \frac{\eta^2 (\Delta_c- NU_0/2)}{(\Delta_c - NU_0/2)^2 + \kappa^2}.
\label{eqn:int_strength}
\end{equation}
The many-body Hamiltonian for the BEC then reduces to $\tilde{H} = -\int d\rr \Psi^\dagger(\rr)\frac{\hbar^2 \nabla^2}{2m}\Psi(\rr) + H_\mathrm{int}^\mathrm{cav} + H_\mathrm{int}^\mathrm{Ryd}$. The mean-field equation of motion given in~\eref{eqn:elim_GPE} of the main text is found by calculating the Heisenberg equation of motion for $\Psi(\rr,t)$ and again performing the mean-field approximation.
\begin{figure*}
    \centering
    \includegraphics[width =0.8\textwidth]{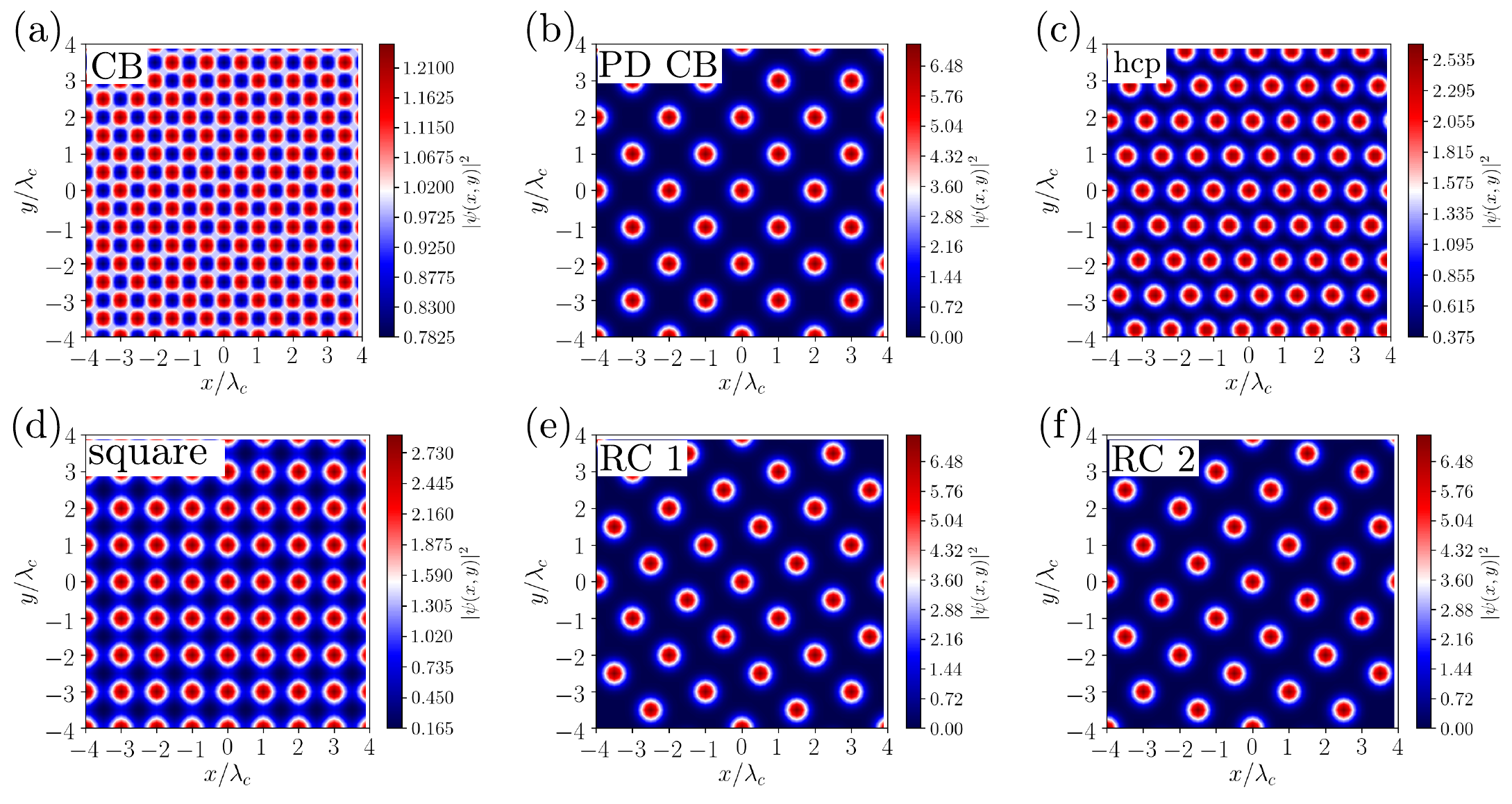}
    \caption{The six non-trivial initial conditions (we refrain from showing the homogeneous state) based on the intuitive picture drawn in~\fref{fig:comb_pot}. (a) checkerboard pattern (CB), (b) period-doubled checkerboard pattern (PD CB), (c) hexagonal closed pack (hcp), (d) square lattice, (e) rotated chain version 1 (RC 1) and (f) rotated chain version 2 (RC 2).}
    \label{fig:initial_con}
\end{figure*}

Note that in both GPE equations presented in the main text [\eeref{eqn:GPE_full} and~\eqref{eqn:elim_GPE}] we neglected the potential term $\propto \cos^2(k_c y)$, which arises due to interference of the pump beams [see~\eref{eqn:MB_at}]. This term is irrelevant for understanding the collective instabilities and the main fundamental results presented in this work also hold if this term is included as we checked by including it in the numerics. However, neglecting this term facilitates the interpretation of the results and provides a method to understand the fundamental physics from an intuitive semi-analytical standpoint as it is outlined in section~\ref{sec:roton} of the main text. In fact, the interference between the two pump beams could even be suppressed in real experimental setups. This can either be achieved by choosing a large enough frequency difference between the two pump beams, or by using two counterpropagating beams with orthogonal polarization. In this case, the model in~\eref{eqn:full_sys} becomes exact.

\section{Variational algorithm}\label{app:var_alg}

To find the self-consistent ground state in section~\ref{sec:sc_GS} we employ a variational algorithm. Here we discuss this algorithm in more detail. In particular, we focus on our choice of initial conditions and argue why the chosen initial conditions are good guesses for the final ground states. In~\fref{fig:comb_pot}, exemplary plots for the total interaction potential
\begin{equation}
U_\mathrm{tot}(\rr,\rr') = U_\mathrm{cav}(\rr,\rr') + U_\mathrm{Ryd}(\rr,\rr'),
\label{eqn:comb_int_pot}
\end{equation}
defined as the sum of the two individual interaction potentials given in~\eref{eqn:dressing_pot} and~\eqref{eqn:cav_int_pot} are shown. These potentials provide a good intuition about the anticipated ground states. Hence, we use these potentials as a guideline to define the initial conditions for the imaginary time evolution. We see that for the two special cases (i) ($k_\mathrm{cav} = k_\mathrm{Ryd}$) and (ii) ($k_\mathrm{cav} = k_\mathrm{Ryd}/\sqrt{2}$), the interaction potentials suggest a checkerboard pattern and a square lattice as a ground state which is indeed the self-consistent ground state in these particular cases [\cf~\ffref{fig:PD_checkerboard}(d) and~\fref{fig:PD_square_glass}(e)]. In addition, to those cases we choose the hexagonal closed packed (hcp) lattice as an additional initial condition for all cases because this would be the ground state for a Rydberg-dressed BEC without cavity interactions~\cite{henkel_supersolid_2012}. From~\fref{fig:comb_pot} we see that case (iii) is more complex than the previous two cases. In this case, the ground state could be a checkerboard lattice with larger period than in case (i) [see red dots in~\fref{fig:comb_pot}(c)] or one of the two lattice configurations indicated by the light green and blue circles in~\fref{fig:comb_pot}(c). Based on the above arguments it also becomes clear that only the cases where $k_\mathrm{Ryd}<k_\mathrm{cav}$ results in modified results. If this condition is not fulfilled the length scale imposed by the Rydberg interaction potential is smaller than the cavity interaction potential which only modifies the corresponding interaction potential locally [see~\fref{fig:comb_pot}(a)].

Based on these intuitive arguments we generate the respective initial states $\psi(\rr) = \sum_i \phi_\sigma(\rr-\mathbf{R}_i)$ by superimposing Gaussian wave functions $\phi_\sigma(\rr) = C \exp[-|\rr|^2/(2\sigma^2)]$. The resultant non-trivial initial conditions (we refrain from showing the homogeneous density) are shown in~\fref{fig:initial_con}. We run the variational algorithm discussed in section~\ref{sec:sc_GS} for all these initial conditions for all cases (i)-(iii). We then compare the groundstate energies calculated via~\eref{eqn:E-functional} and the resultant cavity mode amplitude to estimate whether the respective optimization problem is convex or not.

\newpage

\end{document}